\documentclass[aps,twocolumn,superscriptaddress]{revtex4-1}

\usepackage[T1]{fontenc}
\usepackage[utf8]{inputenc}
\usepackage[english]{babel}
\usepackage{amsmath,amsthm,amssymb}
\usepackage{physics}
\usepackage{graphicx}
\usepackage{float}
\usepackage[linkcolor=black,colorlinks=true]{hyperref}
\usepackage[table]{xcolor} 
\usepackage{tabstackengine} 
\usepackage{accents} 

\newcommand{\bs}[1]{\boldsymbol{#1}}
\newcommand{\K}{\boldsymbol{k}}
\newcommand{\Q}{\boldsymbol{q}}
\newcommand{\Peq}{\mathrel{\phantom{=}}}
\newcommand{\ndagger}{\vphantom{\dagger}}
\newcommand{\ncc}{\vphantom{*}}
\newcommand{\ubar}[1]{\underaccent{\bar}{#1}}

\graphicspath{{./images/}}
\allowdisplaybreaks 

\begin{document}

\title{Role of Topology and Symmetry for the Edge Currents of a 2D Superconductor}

\author{Maximilian F. Holst}
\affiliation{Institute for Theoretical Physics, ETH Zurich, Switzerland}
\author{Manfred Sigrist}
\affiliation{Institute for Theoretical Physics, ETH Zurich, Switzerland}
\author{Mark H. Fischer}
\affiliation{Department of Physics, University of Zurich, Switzerland}

\begin{abstract}
The bulk-boundary correspondence guarantees topologically protected edge states in a two-dimensional topological superconductor.
Unlike in topological insulators, these edge states are, however, not connected to a quantized (spin) current as the electron number is not conserved in a Bogolyubov-de Gennes Hamiltonian.
Still, edge currents are in general present.
Here, we use the two-dimensional Rashba system as an example to systematically analyze the effect symmetry reductions have on the order-parameter mixing and the edge properties in a superconductor of Altland-Zirnbauer class DIII (time-reversal-symmetry preserving) and D (time-reversal-symmetry breaking).
In particular, we employ both Ginzburg-Landau and microscopic modeling to analyze the bulk superconducting properties and edge currents appearing in a strip geometry.
We find edge (spin) currents independent of bulk topology and associated topological edge states which evolve continuously even when going through a phase transition into a topological state.
Our findings emphasize the importance of symmetry over topology for the understanding of the non-quantized edge currents.
\end{abstract}

\maketitle


\section{Introduction}

A major reason that topological phases of matter have attracted much attention over the last two decades are the protected boundary and defect modes. 
These modes are of great technological relevance as they allow for dissipationless one-dimensional transport and have potential for topological quantum computation~\cite{nayak:2008, hasan:2010}.
Additionally, surface zero-energy excitations can be used to identify the non-trivial topology of bulk systems~\cite{beenakker:2013}.

For two-dimensional ($2$D) topological insulating phases, the edge modes are associated with a quantized spontaneous charge (spin) current at the edges of a quantum (spin) Hall insulator~\cite{qi:2011}.
For superconductors, on the other hand, the topological response is a quantized thermal (spin) Hall current due to the presence of chiral (helical) edge modes of Bogolyubov quasiparticles~\cite{sato:2017}; the electron number is not conserved, such that charge (spin) currents cannot be quantized even in a topologically non-trivial phase. 
While not quantized, such edge supercurrents are, however, generally expected from a symmetry perspective.

The $2$D Rashba system---a planar electron system with missing in-plane mirror symmetry---has been extensively discussed in the context of topological superconductivity as it naturally hosts both a topological superconducting phase that conserves and one that breaks time-reversal symmetry (TRS). 
For preserved TRS, the system belongs to class DIII in the Altland–Zirnbauer classification~\cite{schnyder:2008} and realizes a helical superconducting phase if the spin-triplet component of the order parameter is dominant~\cite{lu:2008, sato:2009c}. 
This phase features counter-propagating edge modes which carry a spin current~\cite{iniotakis:2007, tanaka:2009}. 
For broken TRS, class D, a chiral superconductor can be realized  with a single chiral mode at its boundary~\cite{sato:2009b, ghosh:2010}.
In both cases, the edge modes are often associated with a finite current (charge or spin) at the sample's edges.

While non-trivial topology guarantees edge modes, the presence of either current at the boundary can be understood purely on symmetry grounds. 
In particular, the lack of a mirror symmetry leads to a mixing of spin-singlet and spin-triplet order parameters~\cite{gorkov:2001, smidman:2017} with breaking TRS adding further order parameters~\cite{fischer:2018}. 
These order parameter components are furthermore deformed close to the system's boundaries, leading to finite currents~\cite{achermann:2014}. 
However, the role of the topological edge states in the occurrence of these edge currents remains unclear.

In this work, we employ a detailed analysis of the consecutive symmetry reductions when first removing the in-plane mirror symmetry and subsequently breaking TRS. 
In particular, we discuss how the symmetry reductions lead to a mixing of order parameters and allow for edge currents purely on symmetry grounds. 
These edge states show a continuous behavior when changing system parameters, even when going through a topological phase transition.
Moreover, their presence is independent of the presence of protected edge states.

The rest of this paper is organized as follows: 
In Sec.~\ref{sec:bulk}, we discuss the bulk superconducting phases according to the different symmetry reductions.
We discuss the mixing of order parameters within both a Ginzburg-Landau (GL) description and a microscopic tight-binding model.
In the latter case, we solve the gap equations self-consistently for attractive on-site and nearest-neighbor interactions.
In Sec.~\ref{sec:strip}, we discuss the spin and charge edge currents by using the gap functions found in the infinite system within a Bogolyubov-de Gennes (BdG) description of a finite system with strip geometry and also explain their appearance qualitatively within the GL formalism.
We finish by concluding that the observed spin and charge currents vary continuously along the phase transitions between the topologically trivial, helical (class DIII) and chiral (class D) phases and appear due to the symmetry reductions of inversion and TRS rather than due to the topological properties of the corresponding phases.


\section{Bulk Superconductivity}
\label{sec:bulk}


\subsection{Ginzburg-Landau Theory}
\label{sec:bulk:gl}

We start here with a discussion of the symmetry aspects based on a GL formulation. 
First, we consider a square-lattice system within the $xy$ plane which is invariant under symmetry operations of the tetragonal point group $D_{4h}$.
This point group contains spatial inversion $\mathcal{I}$. 
In addition, we assume TRS $\mathcal{T}$. 
A superconducting pairing state, such as the conventional even-parity spin-singlet state transforming according to the trivial irreducible representation (irrep) $A_{1g}$ of $D_{4h}$, can be described through its order parameter $\eta_s$ with the GL free energy density
\begin{equation}
    f = a_{s}(T) \abs{\eta_s}^2 + b_s \abs{\eta_s}^4 ,
    \label{eq:gl_s}
\end{equation}
where we restrict our discussion to the uniform phase deep in the bulk of the material.
The phase transition occurs when the second-order coefficient changes sign, defining the critical temperature $T_{c, s}$ by $a_s(T_{c, s}) = 0$.
Furthermore, $b_s > 0$ is required for the overall stability. 

In this section, we discuss---from a symmetry-based perspective---the implications of successive symmetry reductions by removing first inversion $\mathcal{I}$ and then TRS $\mathcal{T}$. 
For this purpose, we consider a breaking of the mirror symmetry $z\to-z$ which in our microscopic description appears as a spin-orbit coupling of Rashba type.
In a second step, we also add an out-of-plane polarization of the electron spins.


\subsubsection{Inversion-Symmetry Breaking: $C_{4v}$}
\label{sec:bulk:gl:inversion}

Here, we analyze the effect of inversion-symmetry breaking by removing the basal-plane mirror symmetry which reduces the point group from $D_{4h}$ to $C_{4v}$. 
In $C_{4v}$, the two $D_{4h}$-irreps $A_{1g}$ and $A_{2u}$ correspond to the trivial irrep $A_1$. 
We account for this symmetry reduction by extending the order-parameter space, adding $\eta_p$ for an odd-parity spin-triplet pairing state belonging to $A_{2u}$ in the original point group. 
The symmetry reduction leads to an extension of the free energy density, whereby the second-order terms are most relevant to describe the superconducting instability,
\begin{equation}
	f^{\text{R}} = a_s(T) \abs{\eta_s}^2 + a_p(T) \abs{\eta_p}^2 + \epsilon a_{s p}^{\epsilon} \left( \eta_s^* \eta_p^{\ncc}  + \eta_s^{\ncc} \eta_p^* \right) .
\label{eq:gl_sp}
\end{equation}
Here, $a_p(T_{c, p})=0$ defines the transition temperature of the spin-triplet state and 
the symmetry-lowering is represented by the parameter $\epsilon$ which transforms according to $A_{2u}$ and couples the two order-parameter components with strength $a_{s p}^{\epsilon}$.
Thus, the breaking of inversion symmetry causes a mixing of the spin-singlet ($\eta_s$, $s$-wave) and spin-triplet ($\eta_p$, $p$-wave) pairing channels. 

The superconducting instability condition follows from linearizing the GL equations derived from Eq.~\eqref{eq:gl_sp} for the order-parameter components and is given by the highest temperature such that
\begin{equation}
    \det \begin{pmatrix}
        a_s(T) & \epsilon a_{s p}^{\epsilon} \\
        \epsilon a_{s p}^{\epsilon} & a_p(T)
    \end{pmatrix} = 0.
    \label{eq:lin_gleq}
\end{equation}
The off-diagonal elements introduced by the mirror-symmetry breaking raise the transition temperature above that of the bare ones, $T_c > T_{c, s}, T_{c, p}$.
The relative strength of the order-parameter components is determined by the eigenvector of the $2 \times 2$-matrix in Eq.~\eqref{eq:lin_gleq} with zero eigenvalue, such that
\begin{equation}
    \frac{\eta_s}{\eta_p}
    = - \frac{a_p(T)}{\epsilon a_{s p}^\epsilon}.
    \label{eq:bulk_op}
\end{equation}
The evolution of the two order-parameter components for $T < T_c$ is determined by the full GL equations (see App.~\ref{app:glf} for the higher-order terms in the GL free energy functional) and can be chosen such that $\eta_s \in \mathbb{R}$.
The relative phase between $\eta_s$ and $\eta_p$ is either $0$ or $\pi$ depending on the sign of $\epsilon a_{s p}^{\epsilon}$ [Eq.~\eqref{eq:bulk_op}].


\subsubsection{Time-Reversal-Symmetry Breaking: $C_{4v}(C_4)$}
\label{sec:bulk:gl:time_reversal}

We further reduce the symmetry of the system by introducing a TRS-breaking field $m$ perpendicular to the $xy$ plane, physically realized by a Zeeman field and transforming according to $A_{2g}$ in $D_{4h}$.
We restrict the effect of this field to the coupling to the spin degrees of freedom, but not to the charge. 
Owing to the spin-orbit coupling, this field does not introduce paramagnetic depairing~\cite{fischer:2018}. 
However, the field reduces the point group even further to the ferromagnetic point group $C_{4v}(C_4)$, where $C_4$ is the subgroup of unitary symmetry elements not combined with time reversal.
Consequently, the $D_{4h}$-irrep $A_{1u}$ ($A_2$ in $C_{4v}$) now also corresponds to the trivial irrep $A$ of $C_4$ together with $A_1$ of $C_{4v}$.
Denoting the order parameter of the additional $p$-wave contribution as $\eta_{p'}$, the lowest-order terms in the free energy density now read
\begin{equation}
    \begin{split}
        f^{\text{RZ}}
        &= a_s(T) \abs{\eta_s}^2 + a_p(T) \abs{\eta_p}^2 + a_{p'}(T) \abs{\eta_{p'}}^2 \\ 
        &\Peq + \epsilon a_{s p}^{\epsilon} \left( \eta_s^* \eta_p^{\ncc}  + \eta_p^* \eta_s^{\ncc} \right) \\
        &\Peq + i m a_{p p'}^m \left( \eta_p^* \eta_{p'}^{\ncc} - \eta_{p'}^* \eta_{p}^{\ncc} \right) \\
        &\Peq + i \epsilon m a_{s p'}^{\epsilon m} \left( \eta_s^* \eta_{p'}^{\ncc} - \eta_{p'}^* \eta_{s}^{\ncc} \right) .
    \end{split}
    \label{eq:gl_spq1}
\end{equation}

Finally, we discuss the relative phases between the order-parameter components. 
Choosing $\eta_s = \abs{\eta_s}$ real and $\eta_{p(p')} = \abs{\eta_{p(p')}}e^{i\varphi_{p(p')}}$, we can rewrite Eq.~\eqref{eq:gl_spq1} as
\begin{equation}
    \begin{split}
        f^{\text{RZ}}
        &= a_s(T) \abs{\eta_s}^2 + a_p(T) \abs{\eta_p}^2 + a_{p'}(T) \abs{\eta_{p'}}^2 \\
        &\Peq + 2 \epsilon a_{s p}^{\epsilon} \abs{\eta_s} \abs{\eta_p} \cos(\varphi_p) \\
        &\Peq + 2 m a_{p p'}^m \abs{\eta_p} \abs{\eta_{p'}} \sin(\varphi_p - \varphi_{p'}) \\
        &\Peq - 2 \epsilon m a_{s p'}^{\epsilon m} \abs{\eta_s} \abs{\eta_{p'}} \sin(\varphi_{p'}) .
    \label{eq:gl_spq2}
    \end{split}
\end{equation}
These terms can be separately minimized with respect to the phases $\varphi_{p(p')}$: $\varphi_p = 0, \pi$ and $\varphi_{p'} = \pm\pi/2$, respectively, depending on the signs of the coefficients $\epsilon a_{s p}^{\epsilon}$, $m a_{p p'}^m$ and $\epsilon m a_{s p'}^{\epsilon m}$.
The magnitudes $\abs{\eta_i}$, $i \in \{ s, p, p' \}$, of the order-parameter components, however, need to be determined by the full GL equations derived from the full free energy functional stated in App.~\ref{app:glf}.

\begin{figure}[!tb]
    \centering
    \includegraphics{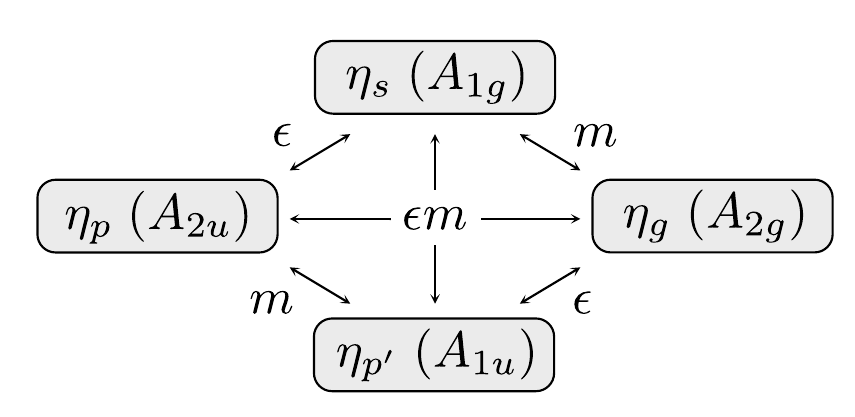}
    \caption{Coupling scheme for the different order-parameter components. 
    In parantheses are the respective irreps of $D_{4h}$.}
    \label{fig:op_coupling_scheme}
\end{figure}

\begin{table}[!tb]
    \centering
    \begin{ruledtabular}
        \begin{tabular}{ccccc|c|c}
            $D_{4h}$ & $\overset{\mathcal{I}}{\longrightarrow}$ & $C_{4v}$ & $\overset{\mathcal{T}}{\longrightarrow}$ & $C_4$ & OP & basis functions \\
            \hline
            $A_{1g}$ & $\rightarrow$ & $A_1$ & $\rightarrow$ & $A$ & $\eta_s$ & $1,\ x^2+y^2$ \\
            $A_{2u}$ & $\rightarrow$ & $A_1$ & $\rightarrow$ & $A$ & $\eta_p$ & $\bs{e}_x y - \bs{e}_y x$ \\
            $A_{1u}$ & $\rightarrow$ & $A_2$ & $\rightarrow$ & $A$ & $\eta_{p'}$ & $\bs{e}_x x + \bs{e}_y y$ \\
            $A_{2g}$ & $\rightarrow$ & $A_2$ & $\rightarrow$ & $A$ & $\eta_g$ & $xy(x^2-y^2)$
        \end{tabular}
    \end{ruledtabular}
    \caption{Mixing of irreps and corresponding order parameters (OP) under consecutive symmetry reductions. 
    The unit vectors $\bs{e}_x$, $\bs{e}_y$ and $\bs{e}_z$ point along the $x$, $y$ and $z$ directions for the superconducting $\bs{d}$ vector, respectively.}
    \label{tab:gl:irrep_mixing}
\end{table}
Table~\ref{tab:gl:irrep_mixing} summarizes the consecutive symmetry reductions and Fig.~\ref{fig:op_coupling_scheme} the coupling of the different order parameters. 
For completeness, we included the $D_{4h}$-irrep $A_{2g}$ ($A_2$ in $C_{4v}$) which, without inversion or TRS, also corresponds to the trivial representation of $C_4$.
A respective order parameter $\eta_g$ couples through $\epsilon$ to $\eta_{p'}$, through $m$ to $\eta_s$ and through both to $\eta_p$.
This order-parameter component has $g$-wave symmetry and within a lattice approach occurs only with pairing interactions beyond nearest-neighbor and next-nearest-neighbor coupling. 
Therefore, we neglect this contribution here and in our microscopic discussions.

\subsection{Microscopic Theory}
\label{sec:bulk:micro}

\begin{figure}[tb]
    \centering
    \includegraphics{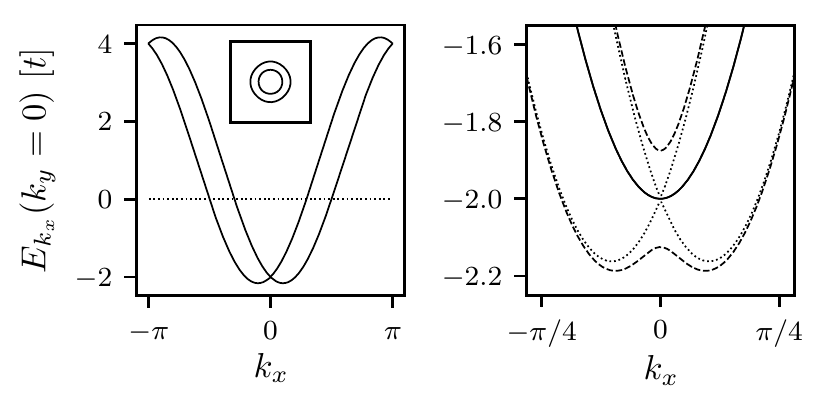}
    \caption{(Left) Band structure of the Rasbha model studied with the dispersion along a high-symmetry axis. 
    We have used $t = 1$, $t' = 0.25$, $\alpha = \alpha' = 0.5$ and $\mu = -3$. 
    The inset shows the corresponding Fermi surfaces. 
    (Right) Zoom of the band-crossing region with the same parameters apart from $\alpha = \alpha' = h = 0$ (solid line), only $h = 0$ (dotted line), and $h = 0.125$ (dashed line).}
    \label{fig:band_structure}
\end{figure}

To illustrate the occurrence of the various phases as discussed in Sec.~\ref{sec:bulk:gl}, we introduce a microscopic model for a $2$D Rashba system and study the superconducting instabilities.
In particular, we consider electrons on a square lattice (lattice constant $a=1$) including Rashba spin-orbit coupling described by the Hamiltonian 
\begin{equation}
	\mathcal{H}_0
	= \sum_{\K, s} \xi_{\K}^{\ndagger} c_{\K, s}^{\dagger} c_{\K, s}^{\ndagger} + \sum_{\K} \sum_{s, s'} \bs{g}_{\K}^{\ndagger} \cdot c_{\K, s}^{\dagger} \bs{\sigma}_{s s'}^{\ndagger} c_{\K, s'}^{\ndagger}.
\label{eq:hsp}
\end{equation}
Here, $c^{\dagger}_{\K, s}$ ($c^{\ndagger}_{\K, s}$) creates (annihilates) an electron with momentum $\K$ and spin $s$, and $\bs{\hat{\sigma}} = \left( \hat{\sigma}^x, \hat{\sigma}^y, \hat{\sigma}^z \right)$ contains the Pauli matrices. 
The dispersion of this Hamiltonian reads
\begin{equation}
    \xi_{\K \lambda} = \xi_{\K} + \lambda \abs{\bs{g}_{\K}},
\end{equation}
where $\lambda=\pm 1$ is the band index for the two spin-split Fermi surfaces.

The first, spin-independent term in Eq.~\eqref{eq:hsp} is given by 
\begin{equation}
    \xi_{\K} = - 2t ( \cos k_x + \cos k_y ) - 4t' \cos k_x \cos k_y - \mu,
    \label{eq:kinetic}
\end{equation}
where $t$ and $t'$ are the  nearest-neighbor (nn) and next-nearest-neighbor (nnn) hopping strengths, respectively, and $\mu$ is the chemical potential.
This spin-independent term respects the square-lattice symmetry $D_{4h}$. The second, spin-dependent, term in Eq.~\eqref{eq:hsp} has the form
\begin{equation}
\begin{split}
    g_{\K}^x &= - \alpha \sin k_y - \alpha' \sin k_y \cos k_x \\
	g_{\K}^y &= + \alpha \sin k_x + \alpha' \sin k_x \cos k_y \\
	g_{\K}^z &= h.
\end{split}
\label{eq:soc}
\end{equation}
The $x$ and $y$ components include nn and nnn Rashba-type spin-orbit couplings of strength $\alpha$ and $\alpha'$. 
Note that this spin-orbit coupling term, which is odd under in-plane mirror symmetry, is the microscopic manifestation of the broken inversion symmetry and hence lowers the system's symmetry from $D_{4h}$ to $C_{4v}$.
Finally, the breaking of TRS is introduced through a Zeeman coupling of an external magnetic field $h=\mu_{\rm B} H^z$ along the $z$ axis. 
In addition to TRS, this field also breaks the mirrors parallel to $z$ and thus further reduces the symmetry from $C_{4v}$ to $C_{4v}(C_4)$ which contains the subgroup of unitary operations $C_4$ as well as the antiunitary operations $C_{4v}\setminus C_4$ that need to be combined with TRS.

For numerical calculations, we use the parameters $t = 1$ (from now on as energy unit), $t' = 0.25$, $\alpha =\alpha' = 0.5$ in the remainder of the paper. 
This results in the Fermi surface and band structure depicted in Fig.~\ref{fig:band_structure}.

To study superconductivity in this model, we introduce the interaction Hamiltonian
\begin{equation}
\begin{split}
    \mathcal{H}_{\text{int}}
    &= U \sum_i n_{i, \uparrow} n_{i, \downarrow} + V \sum_{\langle i, j \rangle} \sum_{s, s'} n_{i, s} n_{j, s'} \\
    	&\Peq + W \sum_{\langle i, j \rangle} \bs{S}_i \cdot \bs{S}_j
\end{split}
\label{eq:hint}
\end{equation}
consisting of an on-site density-density interaction $U$, nn density-density interaction $V$ and nn spin-spin interaction $W$. 
These interactions allow us to investigate the (fully-gapped) phases of the expected phase diagram~\cite{greco:2018, wolf:2020} as we will see below.

In the following, we require $U, V, W < 0$.
In Eq.~\eqref{eq:hint}, we have further introduced the (real-space) electron density $n_{i, s}^{\ndagger} = c_{i, s}^{\dagger} c_{i, s}^{\ndagger}$ and spin density $\bs{S}_i = \sum_{s,s'} c_{i, s}^{\dagger} \bs{\sigma}_{s s'}^{\ndagger} c_{i, s'}^{\ndagger}$, where the operators for electrons at the position $\bs{R}_i$ are defined as
\begin{equation}
    c_{i, s} = \frac{1}{\sqrt{N}} \sum_{\K} c_{\K, s} e^{i \K \cdot \bs{R}_i}
\end{equation}
with $N=N_xN_y$ the total number of lattice sites and assuming periodic boundary conditions.


\subsubsection{Inversion-Symmetry Breaking: $C_{4v}$}
\label{sec:bulk:micro:inversion}

We rewrite the interaction term Eq.~\eqref{eq:hint} only keeping Cooper-pair-scattering processes
\begin{equation}
    \mathcal{H}_{\text{int}} 
    = \frac{1}{2N} \sum_{\K, \K'} \sum_{\{s_i\}} V_{s_1 s_2}^{s_3 s_4} (\K, \K') c_{\K, s_1}^{\dagger} c_{-\K, s_2}^{\dagger} c_{-\K', s_3}^{\ndagger} c_{\K', s_4}^{\ndagger}
\label{eq:hintcooper}
\end{equation}
and decompose the momentum dependence
\begin{equation}
\begin{split}
    V_{s_1 s_2}^{s_3 s_4} (\K, \K')
    &= \sum_{\Gamma, m, \nu, \nu'} u_{\Gamma, m} \left[ \Psi_{\Gamma, m}^{\nu} (\K) (i \hat{\sigma}^{\nu} \hat{\sigma}^y)_{s_1 s_2} \right] \\
    &\Peq \times \left[ \Psi_{\Gamma,m}^{\nu'\ *} (\K') (i \hat{\sigma}^{\nu'} \hat{\sigma}^y)_{s_3 s_4}^{\dagger} \right].
\end{split}
\label{eq:vint}
\end{equation}
Here, $\Psi_{\Gamma, m}^{\nu}$ is the ($m$-th) basis function of the irrep $\Gamma$, see App.~\ref{app:bfd}.
The basis functions are of the structure $\Psi^0 = \psi (\K)$ or $\Psi^i = d^i (\K)$ with $i = x, y, z$ corresponding to spin-singlet and spin-triplet contributions, respectively.

As discussed above, the reduction of the point group symmetry from $D_{4h}$ to $C_{4v}$ due to the mirror-symmetry breaking mixes irreps of $D_{4h}$, see Tab.~\ref{tab:gl:irrep_mixing}.
In the $A_1$ pairing channel, we thus find the basis functions and pairing strengths as given in Tab.~\ref{tab:micro:basis}.
For simplicity, we focus in the following on the interplay of $s$-wave and $p$-wave gap functions but suppress the extended-$s$-wave gap function by choosing $V - 3W = 0$ \footnote{%
The extended-$s$-wave gap function is also suppressed if the chemical potential is tuned such that the Fermi surface is close to the line nodes of the extended $s$-wave gap function. 
In order to have the chemical potential as a tuning parameter for the topological transition in the TRS-broken phase, we chose to fix $V - 3W = 0$.}.
This reduces the number of independent interaction parameters to $U_s = U/2$ and $U_p = V + W$.

\begin{table}[tb]
\centering
\begin{ruledtabular}
\begin{tabular}{cccc}
    & Channel & Basis function & $u_{A_1}$ ($u_A$) \\
    \hline
    $C_{4v}$ & $s$-wave & $\psi_s = 1$ & $U/2$ \\
    & ext.~$s$-wave & $\psi_{s^*} = \cos k_x + \cos k_y$ & $V - 3W$ \\
    & $p$-wave & $\bs{d}_p = \bs{e}_x \sin k_y - \bs{e}_y \sin k_x$ & $V + W$\vspace{2pt}\\
    \hline
    $C_{4v}(C_4)$ & $p'$-wave & $\bs{d}_{p'} = \bs{e}_x \sin k_x + \bs{e}_y \sin k_y$ & $V + W$ \rule{0pt}{2.6ex}
\end{tabular}
\end{ruledtabular}
\caption{Gap function contributions to the trivial irrep and interaction parameters for lacking inversion symmetry only, $C_{4v}$ ($\leftrightarrow u_{A_1}$), and for additional TRS breaking, $C_{4v}(C_4)$ ($\leftrightarrow u_A$).}
\label{tab:micro:basis}
\end{table}

With the help of the decomposition in Eq.~\eqref{eq:hintcooper} and Eq.~\eqref{eq:vint}, we can decouple the full Hamiltonian within a mean-field approximation
\begin{equation}
\begin{split}
    \mathcal{H}_{\text{MF}}
    &= \sum_{\K} \sum_{s, s'} c_{\K, s}^{\dagger} H_{0, s s'}^{\ndagger}(\K) c_{\K, s'}^{\ndagger} \\
    &\Peq + \frac{1}{2} \sum_{\K} \sum_{s, s'} \left( c_{\K, s}^{\dagger} \Delta_{s s'}^{\ndagger}(\K) c_{-\K, s'}^{\dagger} + h.c. \right),
\label{eq:hmf}
\end{split}
\end{equation}
where $\hat{H}_{0}(\K) = (\xi_{\K} \hat{\sigma}^0 + \bs{g}_{\K} \cdot \bs{\hat{\sigma}})$ and
\begin{equation}
    \hat{\Delta} (\K)
    = \left[ \Delta_s \psi_s(\K) \hat{\sigma}_0 + \Delta_p \bs{d}_p(\K) \cdot \bs{\hat{\sigma}} \right] (i \hat{\sigma}^y).
\label{eq:inversion:gap}
\end{equation}
To determine the coefficients $\Delta_s$ and $\Delta_p$, we solve the self-consistent gap equation 
\begin{equation}
    \Delta_{s s'}^{\ndagger}(\K) 
    = -\frac{1}{\beta N}\sum_{n,\K'}\sum_{s_3, s_4}V_{s s'}^{s_3 s_4}(\K, \K') F_{s_4 s_3}(\K', \omega_n),
\label{eq:scge}
\end{equation}
where $F_{s s'}(\K, \omega_n)$ is the anomalous Green's function and $\omega_n = (2n + 1)\pi k_{\rm B} T$ are the fermionic Matsubara frequencies. 
We can express $\hat{F}(\K, \omega_n)$ as
\begin{equation}
\begin{split}
    \hat{F} 
    &= \hat{G}_0(\K, \omega_n) \hat{\Delta}(\K) \\ 
    &\Peq \times \left[ \hat{G}_0(\K, \omega_n)^{-1, T} + \hat{\Delta}(\K)^{\dagger} \hat{G}_0(\K, \omega_n) \hat{\Delta}(\K) \right]^{-1}
\end{split}
\label{eq:fullF}
\end{equation}
with the help of the single-particle Green's function for the non-interacting Hamiltonian Eq.~\eqref{eq:hsp},
\begin{equation}
    \hat{G}_0(\K, \omega_n) = G_+(\K, \omega_n) \hat{\sigma}^0 + G_-(\K, \omega_n) \Big( \frac{\bs{g}_{\K}}{\abs{\bs{g}_{\K}}} \cdot \bs{\hat{\sigma}} \Big)
\label{eq:g0}
\end{equation}
and
\begin{equation}
    G_\pm(\K, \omega_n) = \frac{1}{2}\Big( \frac{1}{i \omega_n - \xi_{\K +}} \pm \frac{1}{i \omega_n - \xi_{\K -}}\Big).
\end{equation}
For details of the calculations, we refer to App.~\ref{app:scge}. 

\begin{figure}[!tb]
    \centering
    \includegraphics{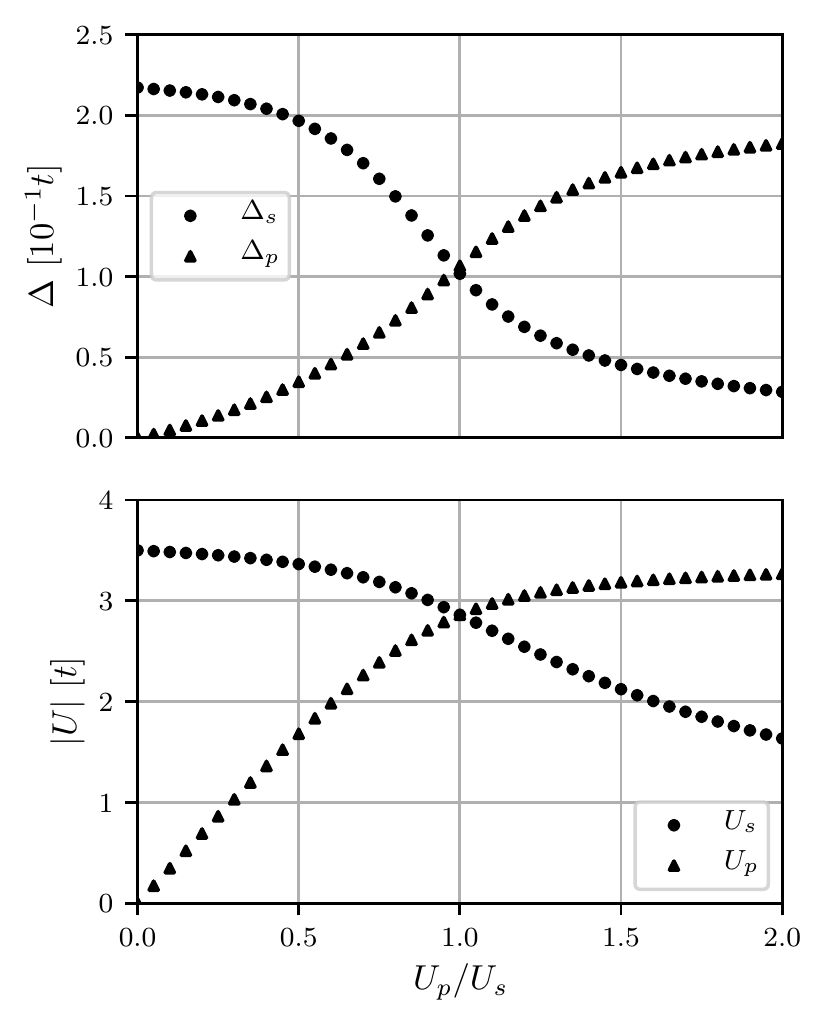}
    \caption{(Top) Order parameter coefficients for different ratios of the interaction strengths at constant $T = 0.5 T_c$. 
    $T_c$ is defined by $U_s = -3.5$, $U_p = 0$. 
    Single-particle parameters as in Fig.~\ref{fig:band_structure}. 
    (Bottom) Interaction strengths for different ratios at constant $T_c$.}
    \label{fig:scge_inversion}
\end{figure}

Figure~\ref{fig:scge_inversion} shows the spin-singlet and spin-triplet contributions to the gap function depending on the interaction strength $U_p / U_s$ (top). 
For better comparison, we choose the absolute strength of the interactions (bottom) such that the transition temperature remains constant.
Due to the lack of inversion symmetry, the singlet- and triplet-contributions are non-vanishing for all $U_p \neq 0$. 
Both coefficients are real and have a relative phase of $0$, in agreement with the results in Sec.~\ref{sec:bulk:gl}.
Depending on the ratio $U_p/U_s$, either the $s$-wave or the $p$-wave component is dominant and the phase-transition temperature is mainly determined by the corresponding interaction parameter.
The transition from dominant $s$- to dominant $p$-wave is a topological phase transition with the latter being a helical superconductor. 
The helical nature arises due to (chiral) $p\pm ip$ pairing for up- and down-spins with opposite chirality.

\begin{figure}[!t]
    \centering
    \includegraphics{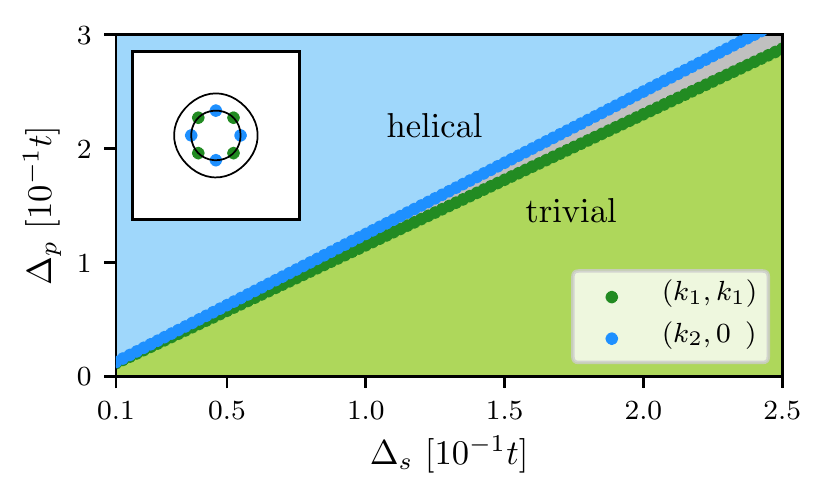}
    \caption{Phase diagram for the inversion-symmetry-broken case with a topologically trivial (green) and a non-trivial, helical phase (blue). 
    Note that there is a finite region (gray) in between, where the gap has point nodes. 
    Inset: Point nodes at the critical ratios corresponding to the trivial and the topological phase.}
    \label{fig:phase_diagram_helical}
\end{figure}

Figure~\ref{fig:phase_diagram_helical} shows the phase diagram as a function of the two gap-function coefficients. 
Note that for an isotropic system with circular Fermi surfaces, this transition occurs at $|\Delta_s|=|\Delta_p|$ \cite{lu:2008, sato:2009} when one of the Fermi surfaces is completely gapless.
On a lattice, however, the topologically trivial and non-trivial phases are separated by a finite region in parameter space where the gap has (point) nodes.

As a final comment, we note that when the extended $s$-wave is not suppressed by a suitable choice of the spin-spin interaction, we find parameter regions where no topological phase transition occurs at all.
Instead, the nearest-neighbor interactions are more beneficial for the extended $s$-wave than for the $p$-wave and consequently, there is a transition from dominant $s$-wave to dominant extended $s$-wave gap function.


\subsubsection{Time-Reversal-Symmetry Breaking: $C_{4v}(C_4)$}
\label{sec:bulk:micro:time_reversal}

When we couple the electron spins to a finite magnetic field in $z$ direction, Eq.~\eqref{eq:soc}, the point group symmetry is further reduced from $C_{4v}$ to $C_{4v}(C_4)$. 
Consequently, an additional $p$-wave component mixes into the above discussed gap function, see Sec.~\ref{sec:bulk:gl} and Tab.~\ref{tab:micro:basis}.
The full gap function has the form
\begin{equation}
    \hat{\Delta} (\K)
    = \big\{ \Delta_s \psi_s(\K) \hat{\sigma}^0
    + [\Delta_p \bs{d}_p(\K) - i\Delta_{p'} \bs{d}_{p'} (\K)] \cdot \bs{\hat{\sigma}} \big\} (i \hat{\sigma}^y),
\label{eq:trs_gap}
\end{equation}
where the factor $-i$ is chosen such that $\Delta_{p'}$ is positive and real.
The coefficients $\Delta_s$, $\Delta_p$ and $\Delta_{p'}$ can again be determined with the help of the self-consistent gap equation [Eq.~\eqref{eq:inversion:gap}].

\begin{figure}[tb]
\centering
\includegraphics{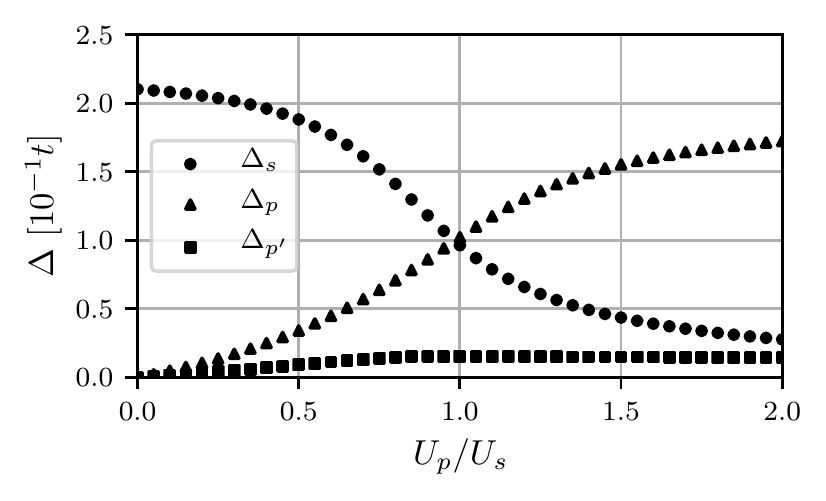}
\caption{Order parameter coefficients for different ratios of the interaction strengths at $T = 0.5 T_c$. $T_c$ is defined by $U_s = -3.5$, $U_p = 0$.}
\label{fig:scge_time_reversal}
\end{figure}

Figure~\ref{fig:scge_time_reversal} shows the resulting gap coefficients using the same parameters as above with an applied field of $h = 0.125$. 
For any finite $U_p$, we observe a small but finite contribution from $\Delta_{p'}$.
This finite contribution leads to an imbalance between the two chiral spin sectors, such that now a finite charge current at the boundary is in principle expected~\cite{fischer:2018}.

Note that unlike the TRS-preserving case, where a topological phase is possible for any chemical potential, in the TRS-broken case, the superconducting phase is only non-trivial, in other words chiral, when
\begin{equation}
    \Delta_s < \sqrt{h^2 - \delta\mu^2},
    \label{eq:chiral_condition}
\end{equation}
where $\delta\mu = \mu + 4(t + t')$ measures the chemical potential from the band crossing in the single-particle spectrum in the absence of a Zeeman magnetic field, see Fig.~\ref{fig:band_structure}~\cite{sato:2009b, ghosh:2010}.

At the topological transition, the bulk gap closes at $\K = 0$, where the mean-field Hamiltonian in the Nambu basis $\mathcal{N}^{\dagger} = \frac{1}{\sqrt{2}} \begin{pmatrix} c_{\K, \uparrow}^{\dagger} & c_{\K, \downarrow}^{\dagger} & c_{-\K, \uparrow}^{\ndagger} & c_{-\K, \downarrow}^{\ndagger} \end{pmatrix}$ reduces to
\begin{equation}
    \ubar{H}_{\text{MF}}(\K =0)
    = \begin{pmatrix}
        -\delta\mu\hat{\sigma}^0 + h\hat{\sigma}^z & \Delta_s(i\hat{\sigma}^y) \\
        -\Delta_s(i\hat{\sigma}^y) & \delta\mu\hat{\sigma}^0 - h\hat{\sigma}^z
    \end{pmatrix}.
\end{equation}
The condition in Eq.~\eqref{eq:chiral_condition} results from the requirement that one of the eigenvalues becomes $0$.
In particular, it follows that the phase transition is purely driven by the strength of the $s$-wave component as, at $\K = 0$, the odd $p$-wave components are always zero.


\section{Strip Geometry and Edge Currents}
\label{sec:strip}


\subsection{Microscopic Current Description}
\label{sec:strip:micro}

We have seen in the discussion of bulk superconductivity how the system can be in two distinct topologically non-trivial phases, a helical one in class DIII and a chiral one in class D. 
Here, we study the edge currents and the effect of topology on their occurrence in a strip geometry, where the system is open in $y$ direction and periodic in $x$ direction. 
Consequently, we can use a momentum representation for the $x$ direction and investigate the system within a BdG approach. 
The resulting Hamiltonian is block diagonal in $k_x$ and can be diagonalized numerically (see App.~\ref{app:bdg}).

For simplicity, along the $y$ direction, we assume spatially constant gap functions determined by their bulk value [Eq.~\eqref{eq:scge}].
This assumption should be valid for a sufficiently wide strip.

Before turning to the strip geometry, we derive expressions for the charge and spin currents in an infinite system. 
For this purpose, we start from the respective densities
\begin{equation}
    n_{i}^{\nu} = \sum_{s, s'} c_{i, s}^{\dagger} \sigma_{s s'}^{\nu \ndagger} c_{i, s'}^{\ndagger},
\end{equation}
where $\nu=0$ corresponds to the charge and $\nu=x,y,z$ to the respective spin components.
Using the Heisenberg equation of motion,
\begin{equation}
    \partial_t^{\ncc} n_{i}^{\nu} = \frac{i}{\hbar} \comm{\mathcal{H}_{0}}{n_{i}^{\nu}},
    \label{eq:strip:heisenberg}
\end{equation}
we derive the current operator $\bs{J}_{\Q}^{\nu}$ from the continuity equation. 
In particular, we write
\begin{equation}
    \partial_t n_{\Q}^{\nu} = - i \Q \cdot \bs{J}_{\Q}^{\nu} + P_{\Q}^{\nu}
    \label{eq:strip:continuity_eq}
\end{equation}
and identify all terms proportional to $\bs{q}$---in other words terms that can be written as a divergence---as current terms $\bs{J}_{\Q}^{\nu}$ and all other terms as precession terms $P_{\Q}^{\nu}$ (see App.~\ref{app:currents}).
These latter terms only appear for the spin currents due to the Rashba spin-orbit coupling~\cite{rashba:2003, vorontsov:2008}: unlike charge, spin is not a conserved quantity in the presence of spin-orbit coupling and thus precesses in an effective (momentum-dependent) magnetic field generated by the orbital angular momentum of the electrons.

Currents in $x$ direction at position $y$ can be expressed through the $x$ component of $\bs{J}_{\Q}^{\nu}$ as
\begin{equation}
    J^{x, \nu}(y) = \frac{1}{\sqrt{N}} \sum_q J_{\Q = (0, q)}^{x, \nu} e^{i q y} ,
    \label{eq:current_operator}
\end{equation}
where $y$ is now a continuous variable. Given the structure of the Hamiltonian Eq.~\eqref{eq:hsp}, the current operators contain contributions living on a site ($y=l$) or a bond between two sites ($y=l\!+\!1/2$),
\begin{align}
    J^{x, \nu}(l) 
    &= \sum_{k_xs, s'} c_{k_x, l, s}^{\dagger} J_{s s'}^{x, \nu, \text{d} \ndagger}(k_x) c_{k_x, l, s'}^{\ndagger} \label{eq:strip:site} \\
    \begin{split}
    J^{x, \nu}\left(l+\frac{1}{2}\right) 
    &= \sum_{k_x,s, s'} \left[ c_{k_x, l+1, s}^{\dagger} J_{s s'}^{x, \nu, \text{od} \ndagger}(k_x) c_{k_x, l, s'}^{\ndagger}  +h.c. \right],
    \end{split}
    \label{eq:strip:bond}
\end{align}
where the superscripts \emph{d} and \emph{od} indicate that the components are diagonal and off-diagonal in the BdG formulation, respectively.
For the charge current, we find
\begin{align}
    \hat{J}^{x, 0, \text{d}}(k_x)
    &= \frac{ 2 t \sin k_x \hat{\sigma}^0 + \alpha \cos k_x \hat{\sigma}^y }{N_x \hbar} \\
    \hat{J}^{x, 0, \text{od}}(k_x)
    &= \frac{ 4 t' \sin k_x \hat{\sigma}^0 + \alpha' \left( \cos k_x \hat{\sigma}^y + i \sin k_x \hat{\sigma}^x \right) }{2N_x \hbar}
\end{align}
and for the spin-$z$ current
\begin{align}
    \hat{J}^{x, z, \text{d}}(k_x)
    &= \frac{ 2 t \sin k_x \hat{\sigma}^z }{N_x \hbar} \\
    \hat{J}^{x, z, \text{od}}(k_x)
    &= \frac{ 2 t' \sin k_x \hat{\sigma}^z }{N_x \hbar}.
\end{align}

In the following, we use these expressions to calculate the ground-state expectation values of the currents,
\begin{equation}
    \mathcal{J}^{x, \nu} (y) = \sum_{E < 0} \sum_{k_x} \expval{J^{x, \nu} (y)}{E, k_x}.
    \label{eq:strip:gs_currents}
\end{equation}
Here, the states $\ket{E, k_x}$ are the eigenstates obtained from the BdG-Hamiltonian for the strip geometry $1<y<N_y$ and the current is evaluated for $y=l$ and $y=l\!+\!1/2$.
For visualization, we distribute the current contributions at $y=l\!+\!1/2$ symmetrically distributed to the adjacent sites at $y=l$ and $y=l+1$, see App.~\ref{app:currents}.


\subsubsection{Helical Superconducting Phase}
\label{sec:strip:micro:helical}

\begin{figure}[!tb]
    \centering
    \includegraphics{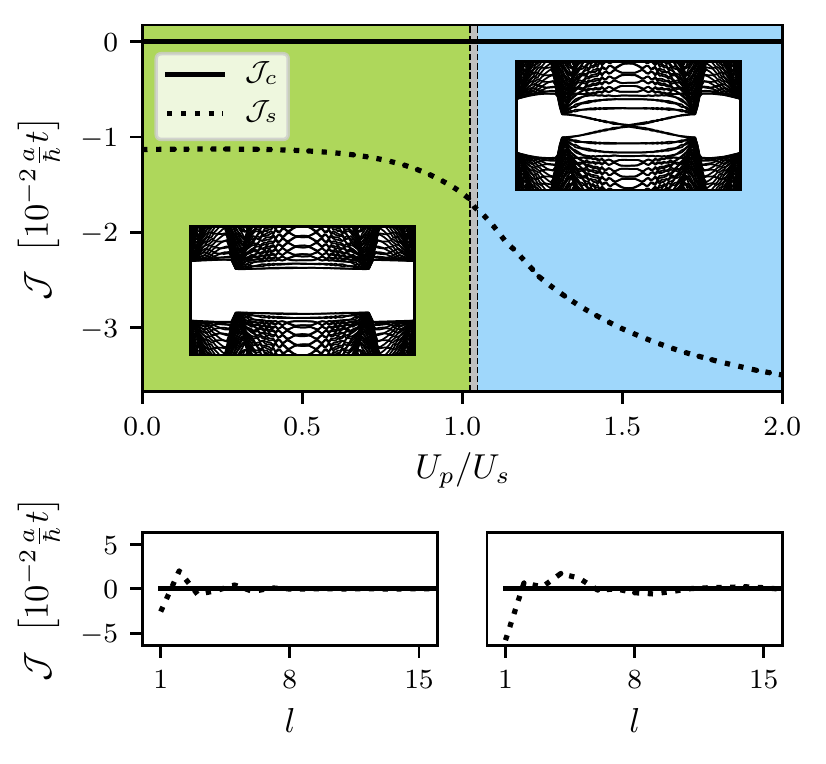}
    \caption{(Top) Integrated current over half of the strip as a function of $U_{p}/U_{s}$. 
    The insets show examples of the spectrum along $k_x$ in the trivial and the topological phase; the latter featuring counter-propagating edge states. 
    (Bottom) Charge and spin currents for the trivial phase (-left) and the helical phase (-right) with parameters as for the corresponding insets.}
    \label{fig:currents_helical_combined}
\end{figure}

After having derived the expressions for the current operators, we can now turn to the strip geometry. 
We start by investigating the situation with TRS, such that the fully-gapped system is either in a trivial or else in a helical superconducting phase. 
As in the infinitely extended system, the gap in the spectrum closes when tuning the system through the transition between the two phases, here by tuning the ratio $U_p/U_s$. 
In the strip geometry, there are additional topological edge states in the helical phase as expected from this topological phase, see insets of Fig.~\ref{fig:currents_helical_combined}. 
These edge states are characterized by counter-propagating spin-currents along the two edges of the strip.

Figure~\ref{fig:currents_helical_combined} shows spin and charge currents both integrated over half the strip, $l = 1, ..., \lfloor N_y/2 \rfloor$, $N_y = 100$, (top) and as a function of position for both phases (bottom). 
As in Fig.~\ref{fig:scge_inversion}, we have chosen the interaction parameters such that the critical temperature stays constant.
In agreement with TRS, the charge currents vanish in the whole parameter range. 
The spin currents, on the other hand, are non-zero both in the trivial and the topological phase. 
Importantly, the integrated value of the spin current changes continuously with the tuning parameter $U_p/U_s$. 
This means, in particular, that the (gapped) bulk of the material---due to the Rashba spin-orbit coupling---carries the observed spin currents while the appearance of topological edge currents does not lead to any discontinuity. 


\subsubsection{Chiral Superconducting Phase}
\label{sec:strip:micro:chiral}

\begin{figure}[tb]
    \centering
    \includegraphics{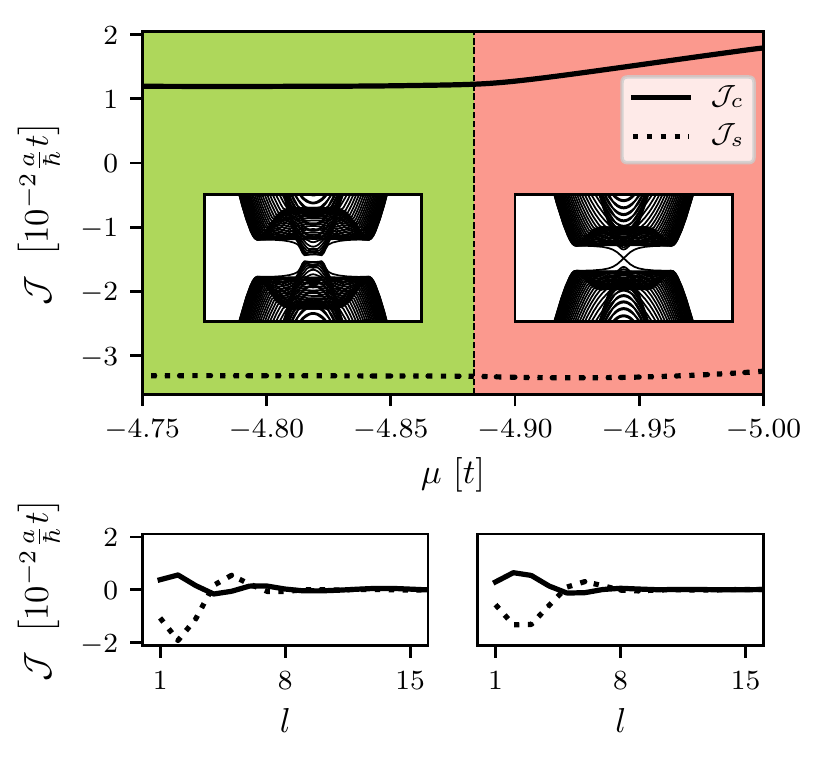}
    \caption{(Top) Integrated current over half of the strip as a function of the chemical potential $\mu$ with $h = 0.125$. 
    The interaction strengths are kept constant at $U_s = -2.5$ and $U_p = -5.0$. 
    The insets show the spectrum for the trivial ($\mu = -4.8$, left) and the topological ($\mu = -4.95$, right) phase along $k_x$; the latter featuring chiral edge states.
    (Bottom) Charge and spin currents for the trivial (-left) and the chiral (-right) phases with parameters as for the corresponding insets.}
\label{fig:currents_chiral_combined}
\end{figure}

Finally, we turn to the TRS broken situation. 
In order to access both the trivial and the chiral phase, we tune the chemical potential $\mu$ at fixed magnetic field strength, see Eq.~\eqref{eq:chiral_condition}. 
Similar to entering the helical phase, the transition to the chiral phase is characterized by a (bulk) gap closing and the subsequent gap opening with additional chiral edge states. 
Examples of the spectra both in the trivial and the chiral phase are shown as insets in Fig.~\ref{fig:currents_chiral_combined}. 
Note that we focus on chemical potentials close to $\mu=-5$ in order to access both the trivial and the topological phases.

Figure~\ref{fig:currents_chiral_combined} summarizes our results for the spin and charge currents in both phases in class D, the TRS broken case. 
Using $U_s = -2.5$ and $U_p = -5.0$, the superconducting phase would be helical in the absence of a magnetic field. 
As a result, the system has a partially spin-polarized charge current along the edge.


\subsection{Ginzburg-Landau Current Description}
\label{sec:strip:gl}

The same surface current behavior as in Sec.~\ref{sec:strip:micro} can also be found qualitatively from a symmetry point of view within the GL formalism.
Generally, besides the GL free energy contributions discussed in Sec.~\ref{sec:bulk:gl}, we need to take into account that the edge of a system affects unconventional order parameters through pair scattering.
As for the microscopic system, we consider an edge with normal vector $\bs{n} = (0, 1, 0)$ parallel to the $y$ axis.
The effect of the surface is described by additional terms in the GL free energy functional which couple the bulk order parameter components $\eta_s, \eta_p, \eta_{p'}$ to the surface normal $\bs{n}$, see App.~\ref{app:glf} \footnote{%
The surface normal vector $\bs{n}$ transforms under $D_{4h}$-point group operations as $E_u\oplus A_{2u}$.}.
These terms determine the boundary conditions and therefore result in a spatial dependence of the order parameters $\eta_i = \eta_i(y)$, $i\in\{s, p, p'\}$, near the edge.
Consequently, additional gradient terms in powers of $\bs{\Pi}\eta_i$ need to be considered in the GL description as well, see App.~\ref{app:glf}.
The covariant gradient $\bs{\Pi} = \bs{\nabla} -i2\pi/\phi_0\bs{A}$ contains the $U(1)$ gauge potential $\bs{A}$ \footnote{%
The magnetic vector potential $\bs{A}$ transforms under $D_{4h}$-point group operations as $E_u\oplus A_{2u}$ and is odd under TRS.}.
Neglecting the magnetic field induced screening currents, the expression for the charge current parallel to the edge obtained from the variation of the free energy with respect to $\bs{A}$ is given by
\begin{equation}
    \begin{split}
        J_{\text{charge}}^x
        &= \frac{4 \pi c}{\Phi_0} \bigl[ K_{p p'}^{\ncc} \sin(\varphi_{p'} - \varphi_p) \partial_y \abs{\eta_p} \abs{\eta_{p'}} \\
        &\Peq - \epsilon K_{s p'}^{\epsilon} \sin(\varphi_{p'}) \partial_y \abs{\eta_s} \abs{\eta_{p'}} \Bigr].
    \end{split}
    \label{eq:gl_charge_current}
\end{equation}
It is important to note that---in agreement with our findings in Sec.~\ref{sec:strip:micro}---the charge current relies on the presence of the $\eta_{p'}$-order-parameter component which appears in the bulk through the spin polarization $m$ (TRS-breaking situation). 
It is connected to the spatial variation of the order parameter near the surface and consequently limited to a range of the coherence length towards the bulk of the superconductor. 
Note again that we have neglected here---as we have done in the microscopic description---the effect of screening of the magnetic field. 
The corresponding screening currents would partially compensate the surface currents on a length scale of the London penetration depth. 

The description of spin-$z$ currents \footnote{%
Due to the presence of spin-orbit coupling, the current is technically an angular-momentum current. 
For simplicity, we will denote the current here nevertheless as a spin current.} requires an additional coupling of the order parameters to the corresponding $SU(2)$ gauge potential $\bs{A}^z$ \footnote{%
The spin-$z$ vector potential $\bs{A}^z$ transforms under $D_{4h}$-point group operations as $E_u\oplus A_{1u} = (E_u\otimes A_{2g})\oplus (A_{2u}\otimes A_{2g})$, where $A_{2g}$ is the irrep under which the spin-$z$ component transforms, and is---as opposed to $\bs{A}$---even under TRS.
}\cite{rebei:2006, jin:2006, berche:2012}.
Restricting our discussion to the TRS case and considering only linear couplings in $\bs{A}^z$, the order parameter components $\eta_s$, $\eta_p$ lead to a spin-$z$ current contribution
\begin{equation}
    \begin{split}
        J_{\text{spin-$z$}}^x
        &\sim \epsilon L_{sp}^{\epsilon}\left[(\partial_y\eta_s)^*\eta_p +c.c.\right] \\
        &\Peq +\epsilon L_{ps}^{\epsilon}\left[(\partial_y\eta_p)^*\eta_s +c.c.\right]
    \end{split}
    \label{eq:gl_spin_current}
\end{equation}
by varying the free energy with respect to $\bs{A}^z$, see App.~\ref{app:glf}.
The second term is likely the dominant contribution as the variation of the $s$-wave component near the surface is typically much smaller than that of the (unconventional) $p$-wave component.
Note that the current is not affected by a sign change of $\epsilon$ because this would be absorbed by the relative sign between the two order-parameter components.

Finally, we would like to state that within the GL formulation, we cannot distinguish between a topologically trivial and non-trivial situation. 
This indicates that both the charge and spin currents are arising as a result of the order parameter symmetry, but not of topology, as we already observed in the microscopic discussion above. 


\section{Conclusion}

While the existence of protected edge states in an unconventional superconductor is guaranteed by a non-trivial topology, the topological response is connected to heat (energy) transport which is experimentally much harder to access than spontantious supercurrents at the boundaries. 
In contrast charge or spin supercurrents, however, are not quantized. 
In our work, we have studied the role of symmetry breaking and topology for both bulk order-parameter mixing and edge phenomena for the prototypical 2D Rashba superconductor. 
In particular, we have studied a two-dimensional system with broken in-plane mirror symmetry (class DIII) and further TRS breaking (class D).
Interestingly, the appearance of topologically protected edge states has no singular effect on the edge currents in a strip geometry, in other words, the currents change continuously even when entering the topologically non-trivial phases. 
Qualitatively, the edge behavior can be understood also within a phenomenological GL treatment which is not sensitive to topological phase transitions.
In particular, both spin and charge currents can be well described through the coupling of order parameters close to the system's boundaries using group theoretical arguments. 
An interesting open question is whether this non-singular behavior survives disorder at the boundary.


\begin{acknowledgments}
We thank D.F. Agterberg, E. Arahata, H.B. Braun and T. Neupert for helpful discussions. 
This work has been supported by the Swiss National Science Foundation (SNSF) through Division II (No. 184739).
\end{acknowledgments}

\bibliographystyle{apsrev4-2}
\bibliography{main}

\clearpage
\appendix
\section{Ginzburg-Landau Free Energy Functional}
\label{app:glf}

The GL free energy functional for a superconductor is an expansion of the free energy in the superconducting order parameters in a form invariant under all symmetry operations of the system. 
The basic symmetries are $U(1)$ gauge, time-reversal and point group symmetries. 
The symmetry reductions we are concerned with are the removal of mirror symmetry, $z \to -z$, represented by the parameter $\epsilon$---microscopically manifested as a Rashba-type spin-orbit coupling---and TRS which we introduce through a (spin) magnetization $m$ along the $z$ axis. 
These two parameters allow us to include symmetry reducing terms into the GL free energy which couple order parameters of different pairing channels with regard to the original symmetry group. 
It is sufficient to introduce $\epsilon$ and $m$ in the terms second order in the superconducting order parameters to cover their main effect. 
Furthermore, we orient the system to the $xy$ plane and neglect in-plane magnetic fields as they are irrelevant for our discussion in Sec.~\ref{sec:bulk:gl}.

First, we consider the GL free energy for the case when inversion symmetry is broken ($\epsilon \neq 0$, Sec.~\ref{sec:bulk:gl:inversion}), 
\begin{widetext}
\begin{equation}
    \begin{split}
    	F^{\text{R}} 
    	= \int d^2r \Biggl[ 
    	& \sum_j a_j(T) \abs{\eta_i}^2 + \epsilon a_{s p}^{\epsilon} \left( \eta_s^* \eta_p^{\ncc}  + c.c.  \right) 
    	+ \sum_j b_{j} \abs{\eta_j}^4 + b_{1 s p} \abs{\eta_s}^2 \abs{\eta_p}^2 + b_{2 s p} \left( {\eta_s^*}^2 \eta_p^{2} + c.c. \right) \\
    	& + \sum_j K_{j} \left( \abs{\Pi_x \eta_j}^2 + \abs{\Pi_y \eta_j}^2 \right)
    	+ \epsilon K_{s p}^{\epsilon} \left( \left( \Pi_x \eta_s \right)^* \left( \Pi_x \eta_p \right) + \left( \Pi_y \eta_s \right)^* \left( \Pi_y \eta_p \right) + c.c. \right) 
    	+ \frac{\bs{B}^2}{8 \pi} \Biggr],
    \end{split}
    \label{eq:app:glf:fr}
\end{equation}
\end{widetext}
where $j \in \{ s, p \}$ labels the respective order parameter component, $a_j (T) = a_j' (T - T_{c, j})$, $a_j',b_j, b_{1 s p}, b_{2 s p}, K_j > 0$, and $\bs{\Pi} = \bs{\nabla} - i2\pi/\phi_0\bs{A}$, with $\phi_0$ being the magnetic flux quantum. 
Note that we only included terms which are relevant for our discussion.

With the additional breaking of TRS ($m\neq 0$, Sec.~\ref{sec:bulk:gl:time_reversal}), we introduce $m$ which leads to additional terms in the GL free energy,
\begin{widetext}
\begin{equation}
    \begin{split}
    	F^{\text{RZ}} 
    	= \int d^2r \Biggl[
    	& \sum_j a_j(T) \abs{\eta_j}^2 + \epsilon a_{s p}^{\epsilon} \left( \eta_s^* \eta_p^{\ncc}  + c.c. \right) + i m a_{p p'}^m \left( \eta_p^* \eta_{p'}^{\ncc} - c.c. \right) + i \epsilon m a_{s p'}^{\epsilon m} \left( \eta_s^* \eta_{p'}^{\ncc} - c.c. \right) \\
    	& + \sum_j b_{j} \abs{\eta_j}^4 + \sum_{j < j'} \left\{ b_{1 j j'} \abs{\eta_j}^2 \abs{\eta_{j'}}^2 +  b_{2 j j'} \left( {\eta_j^*}^2 \eta_{j'}^{2} + c.c. \right) \right\} \\
    	& + \sum_j K_{j} \left( \abs{\Pi_x \eta_j}^2 + \abs{\Pi_y \eta_j}^2 \right) + i K_{p p'} b_z \left( \eta_p^* \eta_{p'}^{\ncc} - c.c. \right) \\
    	& + \epsilon K_{s p}^{\epsilon} \left( \left( \Pi_x \eta_s \right)^* \left( \Pi_x \eta_p \right) + \left( \Pi_y \eta_s \right)^* \left( \Pi_y \eta_p \right) + c.c. \right)
        + i \epsilon K_{s p'}^{\epsilon} b_z \left( \eta_{p'}^* \eta_s^{\ncc} - c.c. \right) \\
    	&  + i m K_{p p'}^m \left( \left( \Pi_x \eta_p \right)^* \left( \Pi_y \eta_{p'} \right) + \left( \Pi_x \eta_{p'} \right)^* \left( \Pi_y \eta_p \right) - c.c. \right) + \frac{\bs{B}^2}{8 \pi} \Biggr],
	\end{split}
\end{equation}
\end{widetext}
where $j, j' \in \{ s, p, p' \}$ label the respective order parameter components and $b_z = 2\pi/\phi_0 (\bs{\nabla} \times \bs{A})_z$.

Boundary conditions can be formulated through additional surface terms which are derived as invariant combinations of the order parameter and the surface normal vector $\bs{n}$, 
\begin{equation}
    \begin{split}
	F^{\text{SF}}
	= \int dr \Biggl[ 
	&\sum_j \left\{ g_{1 j} \left( n_x^2 + n_y^2 \right)+ g_{2 j} n_z^2 \right\} \abs{\eta_j}^2 \\
	& + g_3 n_z \left( \eta_s^* \eta_p^{\ncc} + c.c. \right) \Biggr],
	\end{split}
\end{equation}
which takes surface scattering into account. 
The coefficients for $j = s$ vanish usually. 
Note that in the combination with the other order parameters also $\eta_s$ will be affected near the surface.

The charge surface currents are determined by the gradient terms appearing in the expansion of the GL free energy.
To be more precise, we consider the GL equation derived from the variation of the free energy $F^{\text{RZ}} =\int d^2r f^{\text{RZ}}$ w.r.t.~the magnetic vector potential $\bs{A}$ ($U(1)$ gauge potential)
\begin{equation}
    0 = \frac{\partial f^{\text{RZ}}}{\partial\bs{A}} +\bs{\nabla}\times\frac{\partial f^{\text{RZ}}}{\partial (\bs{\nabla}\times\bs{A})}.
\end{equation}
By making use of Ampere's law, $\bs{\nabla}\times\bs{B} = 4\pi/c\bs{J}_{\text{charge}}$, we find that the charge currents are given by
\begin{equation}
    \bs{J}_{\text{charge}} = -c\frac{\partial f^{\text{RZ}}}{\partial\bs{A}} -c\bs{\nabla}\times\frac{\partial\left(f^{\text{RZ}} -\frac{\bs{B}^2}{8\pi}\right)}{\partial(\bs{\nabla}\times\bs{A})}.
\end{equation}
Assuming a surface normal $\bs{n} = (0, 1, 0)$ and therefore a spatial $y$ dependence of the order parameter, $\eta_i = \eta_i(y)$, and neglecting the screening currents, $\bs{A} \to 0$, yields Eq.~(\ref{eq:gl_charge_current}).

The description of spin-$z$ surface currents within the GL formalism requires the introduction of the corresponding $SU(2)$ gauge potential $\bs{A}^z = (A_x^z, A_y^z)$ transforming according to $E_u = E_u\otimes A_{2g}$ with corresponding basis $\{A_y^z, -A_x^z\}$.
For simplicity, we restrict our discussion to the TRS situation only.
In this case, the additional field gives rise to two new terms in the GL free energy density $f^{\text{R}}$ [Eq.~\eqref{eq:app:glf:fr}], where $F^{\text{R}} = \int d^2r f^{\text{R}}$,
\begin{align}
    &\epsilon L_{sp}^{\epsilon}\left((\Pi_x\eta_s)^*(A_y^z\eta_p) -(\Pi_y\eta_s)^*(A_x^z\eta_p) +c.c.\right),\\
    &\epsilon L_{ps}^{\epsilon}\left((\Pi_x\eta_p)^*(A_y^z\eta_s) -(\Pi_y\eta_p)^*(A_x^z\eta_s) +c.c.\right).
\end{align}
In the same manner as for the charge currents, the spin-$z$ currents are determined by a variation of the free energy functional w.r.t.~$\bs{A}^z$, resulting in Eq.~(\ref{eq:gl_spin_current}).

Finally, we would like to note that a more careful analysis of the surface currents also requires to take into account the effects of the additional symmetry reduction caused by the presence of the edge and the accompanied Cooper pair breaking.

\section{Decomposition of the Interaction Potential into Point Group Basis Functions}
\label{app:bfd}

The interaction potential $V_{s_1 s_2}^{s_3 s_4} (\K, \K')$ arising from the on-site density-density ($U$), nearest-neighbor density-density ($V$) and nearest-neighbor spin-spin interactions ($W$) can be decomposed into the basis functions for the irreps of the underlying point group $D_{4h}$ (inversion and TRS case).
Its general form is then given by
\begin{equation}
    \begin{split}
        V_{s_1 s_2}^{s_3 s_4} (\K, \K')
        &= \sum_{\Gamma, m, \nu, \nu'} u_{\Gamma, m} [ \Psi_{\Gamma, m}^{\nu} (\K) (i \hat{\sigma}^{\nu} \hat{\sigma}^y)_{s_1 s_2} ] \\
        &\Peq\times [ \Psi_{\Gamma, m}^{\nu'} (\K')^* (i \hat{\sigma}^{\nu'} \hat{\sigma}^y)_{s_3 s_4}^{\dagger} ].
    \label{eq:app:bfd:interaction_decomposition}
    \end{split}
\end{equation}
where the spin-singlet basis functions $\Psi^0 = \psi (\K)$ are given in Tab.~\ref{tab:app:bfd:singlet} and the spin-triplet basis functions $\Psi^i = d^i (\K)$, $i = x, y, z$, in Tab.~\ref{tab:app:bfd:triplet}.

\begin{table}[!tb]
    \centering
    \begin{ruledtabular}
        \begin{tabular}{ccc}
            IR & $\psi_{\Gamma, m} (\K)$ & $u_{\Gamma, m}$ \\
            \hline
            $A_{1g}$ & $1$ & $U/2$ \\
             & $\cos k_x + \cos k_y$ & $V - 3W$ \\
            $A_{2g}$ & - & - \\
            $B_{1g}$ & $\cos k_x - \cos k_y$ & $V - 3W$ \\
            $B_{2g}$ & - & - \\
            $E_g$ & - & -
        \end{tabular}
    \end{ruledtabular}
    \caption{Spin-singlet basis functions $\psi (\K)$.}
    \label{tab:app:bfd:singlet}
\end{table}

\begin{table}[!tb]
    \centering
    \begin{ruledtabular}
        \begin{tabular}{ccc}
            IR & $\bs{d}_{\Gamma, m} (\K)$ & $u_{\Gamma, m}$ \\
            \hline
            $A_{1u}$ & $\bs{e}_x \sin k_x + \bs{e}_y \sin k_y$ & $V + W$ \\
            $A_{2u}$ & $\bs{e}_x \sin k_y - \bs{e}_y \sin k_x$ & $V + W$ \\
            $B_{1u}$ & $\bs{e}_x \sin k_x - \bs{e}_y \sin k_y$ & $V + W$ \\
            $B_{2u}$ & $\bs{e}_x \sin k_y + \bs{e}_y \sin k_x$ & $V + W$ \\
            $E_u$ & $\{ \bs{e}_z \sin k_x, \bs{e}_z \sin k_y \}$ & $2(V + W)$
        \end{tabular}
    \end{ruledtabular}
    \caption{Spin-triplet basis functions $\bs{d} (\K)$.}
    \label{tab:app:bfd:triplet}
\end{table}

\section{Derivation of the Self-Consistent Gap Equation}
\label{app:scge}

We follow the general procedure exploited by Ref.~\cite{mineev:1999}. 
The units are chosen such that $\hbar = k_B = 1$.
In a first step, we introduce the Matsubara causal Green's function
\begin{align}
    G_{s s'}^{\ndagger} (\K, \tau)
    &= -\expval{\mathbb{T} c_{\K, s}^{\ndagger} (\tau) c_{\K, s'}^{\dagger}} \label{eq:app:scge:greensfunc_g} \\
    \begin{split}
        &= -\Theta (+\tau) \expval{c_{\K, s}^{\ndagger} (\tau) c_{\K, s'}^{\dagger}} \\
        &\Peq + \Theta (-\tau) \expval{c_{\K, s'}^{\dagger} c_{\K, s}^{\ndagger} (\tau)}
    \end{split}
\end{align}
and the two anomalous Green's functions
\begin{align}
    F_{s s'}^{\ndagger} (\K, \tau)
    &= \expval{\mathbb{T} c_{\K, s}^{\ndagger} (\tau) c_{-\K, s'}^{\ndagger}} \label{eq:app:scge:greensfunc_f}, \\
    F_{s s'}^{\dagger}(\K, \tau)
    &= \expval{\mathbb{T} c_{-\K, s}^{\dagger} (\tau) c_{\K, s'}^{\dagger}}, \label{eq:app:scge:greensfunc_fdag}
\end{align}
where $\mathbb{T}$ is the time-ordering operator, $\Theta$ the Heaviside function and $\tau = it\in\mathbb{R}$.
Making use of the Heisenberg equation---for the full Hamiltonian consisting of the single-particle Hamiltonian [Eq.~\eqref{eq:hsp}] and the interaction Hamiltonian [Eq.~\eqref{eq:hint}]---we can derive the corresponding equations of motion for these Green's functions.
Within a mean-field approximation, we introduce the superconducting order parameter
\begin{equation}
    \Delta_{s_1 s_2}^{\ndagger} (\K)
    = \frac{1}{N} \sum_{\K'} \sum_{s_3, s_4} V_{s_1 s_2}^{s_3 s_4} (\K, \K')F_{s_3 s_4}^{\ndagger} (-\K', \tau = 0)
\label{eq:app:scge:gap}
\end{equation}
and expand the Green's functions in a Fourier series
\begin{align}
    \hat{G} (\K, \tau) 
    &= \frac{1}{\beta} \sum_{n} \hat{G} (\K, \omega_n) e^{-i \omega_n \tau}, \\
    \hat{G} (\K, \omega_n)
    &= \int_0^{\beta} d\tau \hat{G} (\K, \tau) e^{i \omega_n \tau},
\end{align}
where $\omega_n = (2n+1)\pi/\beta$ are the fermionic Matsubara frequencies \cite{mahan:2000}.
Besides $\hat{G}$, we introduce the bare (non-interacting) Green's function
\begin{equation} 
    \hat{G}_0(\K, \omega_n) = \left[ i\omega_n\hat{\sigma}^0 -\hat{H}_0(\K) \right]^{-1}
\end{equation}
with $\hat{H}_0(\K) = \xi_{\K}\hat{\sigma}^0 +\bs{g}_{\K}\cdot\bs{\hat{\sigma}}$, where $\bs{g}_{\K}$ includes both Rashba spin-orbit coupling and a Zeeman field perpendicular to the planar system, both as given in Eq.~\eqref{eq:soc}. 
Corresponding to Eq.~\eqref{eq:g0}, this Green's function can be decomposed into 
\begin{equation}
    \hat{G}_0(\K, \omega_n) = G_+(\K, \omega_n) \hat{\sigma}^0 + G_-(\K, \omega_n) \Big( \frac{\bs{g}_{\K}}{\abs{\bs{g}_{\K}}} \cdot \bs{\hat{\sigma}} \Big)
\label{eq:g0app}
\end{equation}
with the two components
\begin{equation}
    G_\pm(\K, \omega_n) = \frac{1}{2}\Big( \frac{1}{i \omega_n - \xi_{\K +}} \pm \frac{1}{i \omega_n - \xi_{\K -}}\Big).
\end{equation}
and $\xi_{\K \pm} = \xi_{\K}\pm\abs{\bs{g}_{\K}}$.
This can be used to formulate the Gorkov equations
\begin{align}
    \hat{\sigma}^0
    &= \hat{G}_0^{-1}(\K,\omega_n) \hat{G}(\K,\omega_n) + \hat{\Delta} \hat{F}^{\dagger}(\K,\omega_n), \\
    0
    &= \hat{G}_0^{-1}(\K,\omega_n) \hat{F}(\K,\omega_n) - \hat{\Delta} \hat{\ubar{G}}^T(\K,\omega_n), \\
    0
    &= \hat{\ubar{G}}_0^{-1, T}(\K,\omega_n) \hat{F}^{\dagger}(\K,\omega_n) - \hat{\Delta}^{\dagger} \hat{G}(\K,\omega_n),
\end{align}
where the notation $\hat{\ubar{G}}(\K, \omega_n) = \hat{G}(-\K, -\omega_n)$ is used for all Green's functions. 
In the following, we suppress the argument $ (\K, \omega_n )$ and  rearrange these equations to
\begin{align}
    \hat{G}
    &= \left[ \hat{G}_0^{-1} + \hat{\Delta} \hat{\ubar{G}}_0^T \hat{\Delta}^{\dagger} \right]^{-1}, \\
    \hat{F}
    &= \hat{G}_0 \hat{\Delta} \left[ \hat{\ubar{G}}_0^{-1, T} + \hat{\Delta}^{\dagger} \hat{G}_0 \hat{\Delta} \right]^{-1}.
    \label{eq:app:scge:greensfunc_f_gap}
\end{align}
Transforming the superconducting gap [Eq.~\eqref{eq:app:scge:gap}] to the frequency space yields
\begin{align}
    \Delta_{s_1 s_2}^{\ndagger} (\K)
    = -\frac{1}{\beta N} \sum_{n, \K'} \sum_{s_3, s_4} V_{s_1 s_2}^{s_3 s_4} (\K, \K')F_{s_4 s_3}^{\ndagger} (\K', \omega_n).
\end{align}
Using the decomposition of the interaction potential into its basis functions [Eq.~\eqref{eq:app:bfd:interaction_decomposition}] results in
\begin{align}
    \begin{split}
        \hat{\Delta} (\K)
        &= -\frac{1}{\beta N} \sum_{\Gamma, m, \nu} u_{\Gamma, m} \Psi_{\Gamma, m}^{\nu} (\K) (i \hat{\sigma}^{\nu} \hat{\sigma}^y) \\
        &\Peq \times \sum_{\K', \nu'} \Psi_{\Gamma, m}^{\nu'}(\K')^* \\
        &\Peq \times \sum_{n} \Tr{(i \hat{\sigma}^{\nu'} \hat{\sigma}^y)^{\dagger} \hat{F}(\K', \omega_n)}.
        \label{eq:app:scge:gap_omega}
    \end{split}
\end{align}
Making an ansatz for the superconducting gap of the form
\begin{equation}
    \hat{\Delta} (\K) = \sum_{\Gamma, m, \nu} \Delta_{\Gamma, m} \Psi_{\Gamma, m}^{\nu} (\K) (i \hat{\sigma}^{\nu} \hat{\sigma}^y)
\end{equation}
and using the orthogonality of the basis functions $\sum_{\nu} \Psi^{\nu} (i \hat{\sigma}^{\nu} \hat{\sigma}^y)$ with respect to the inner product $\braket{\cdot}{*} = \expval{\frac{1}{2}\Tr{\cdot^{\dagger}*}}_{\K}$, we can project the gap function onto its single components
\begin{equation}
    \begin{split}
        \Delta_{\Gamma, m}
        &= -\frac{u_{\Gamma, m}}{\beta N} \sum_{\K, \nu} \Psi_{\Gamma, m}^{\nu} (\K)^* \\
        &\Peq \times \sum_{n} \Tr{(i \hat{\sigma}^{\nu} \hat{\sigma}^y)^{\dagger} \hat{F}(\K, \omega_n)}.
    \end{split}
\end{equation}
Finally, replacing the anomalous Green's function in the superconducting gap [Eq.~\eqref{eq:app:scge:gap_omega}] by Eq.~\eqref{eq:app:scge:greensfunc_f_gap} results in a self-consistency equation for the gap coefficients $\Delta_{\Gamma, m}$ of the form
\begin{equation}
    \begin{split}
        \Delta_{\Gamma, m}
        &= - \frac{u_{\Gamma, m}}{\beta N} \sum_{\K, \nu} \Psi_{\Gamma, m}^{\nu} (\K)^* \sum_{\Gamma', m'} \Delta_{\Gamma', m'}  \\
        &\Peq \times \sum_{\nu'} \Psi_{\Gamma', m'}^{\nu'} (\K) \mathcal{M}_{\nu \nu'} (\K; \Delta, T)
    \end{split}
\end{equation}
where $\nu, \nu' \in \{ 0, x, y, z \}$ and
\begin{equation}
    \begin{split}
        \mathcal{M}_{\nu \nu'}(\K)
        &= \sum_{n} \Tr \Big\{ (i \hat{\sigma}^{\nu} \hat{\sigma}^y)^{\dagger} \hat{G}_0 (i \hat{\sigma}^{\nu'} \hat{\sigma}^y) \\
        &\Peq \times \left[ \ubar{\hat{G}}_0^{-1, T} + \hat{\Delta}^{\dagger}(\K) \hat{G}_0 \hat{\Delta}(\K) \right]^{-1} \Big\}
    \end{split}
\end{equation}
which is then used to determine the gap coefficients numerically.

\section{Bogolyubov-de Gennes Hamiltonian}
\label{app:bdg}
\begin{figure}[!tb]
    \centering
    \includegraphics{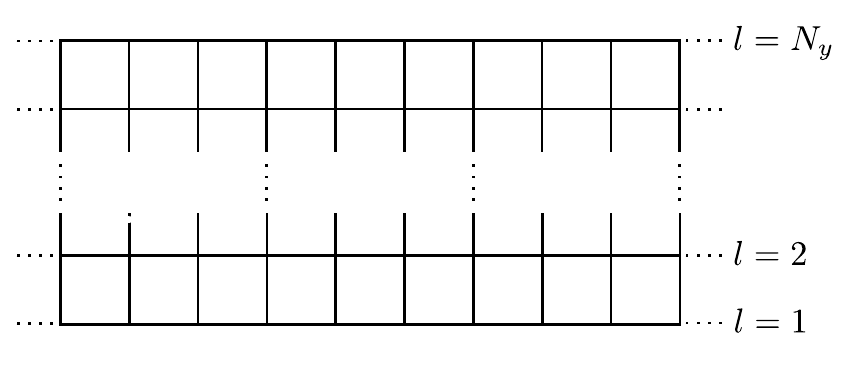}
    \caption{Strip geometry.}
    \label{fig:strip}
\end{figure}
In the Nambu basis
\begin{equation}
    \mathcal{N}_k^{\dagger} 
    = \frac{1}{\sqrt{2}} \begin{pmatrix} \mathcal{C}_{k, 1}^{\dagger} & \cdots & \mathcal{C}_{k, N_y}^{\dagger} & \mathcal{C}_{-k, 1}^{\ndagger} & \cdots & \mathcal{C}_{-k, N_y}^{\ndagger} \end{pmatrix},
\end{equation}
where $\mathcal{C}_{k, l}^{\dagger} = \begin{pmatrix} c_{k, l, \uparrow}^{\dagger} & c_{k, l, \downarrow}^{\dagger} \end{pmatrix}$, the BdG Hamiltonian describing the strip system [Fig.~\ref{fig:strip}] is given by
\begin{equation}
    \mathcal{H}_{\text{BdG}} 
    = \sum_k \mathcal{N}_k^{\dagger} \ubar{H}_{\text{BdG}}^{\ndagger} \mathcal{N}_k^{\ndagger},
\end{equation}
with
\begin{equation}
    \ubar{H}_{\text{BdG}} = \begin{pmatrix} \ubar{H}_0(k) & \ubar{\Delta}(k) \\ \ubar{\Delta}^{\dagger}(k) & -\ubar{H}_0^T(-k) \end{pmatrix},
\end{equation}
up to some constant. 
Note that we use here the short-notation $ k $ for the momentum $ k_x $ along the strip. 
The sub-blocks
\begin{equation}
    \ubar{H}_0(k) = \begin{pmatrix}
        \hat{H}_{\text{d}}^{\ndagger}(k) & \hat{H}_{\text{od}}^{\dagger}(k) & & \\
        \hat{H}_{\text{od}}^{\ndagger}(k) & \ddots & \ddots & \\
        & \ddots &  & \hat{H}_{\text{od}}^{\dagger}(k) \\
        & & \hat{H}_{\text{od}}^{\ndagger}(k) & \hat{H}_{\text{d}}^{\ndagger}(k)
    \end{pmatrix}
\end{equation}
and
\begin{equation}
    \ubar{\Delta}(k) = \begin{pmatrix}
        \hat{\Delta}_{\text{d}}^{\ndagger}(k) & \frac{\hat{\Delta}_{\text{od}}^{\dagger}(k)}{2} & & \\
        -\frac{\hat{\Delta}_{\text{od}}^{T}(-k)}{2} & \ddots & \ddots & \\
        & \ddots &  & \frac{\hat{\Delta}_{\text{od}}^{\dagger}(k)}{2} \\
        & & -\frac{\hat{\Delta}_{\text{od}}^{T}(-k)}{2} & \hat{\Delta}_{\text{d}}^{\ndagger}(k)
    \end{pmatrix}
\end{equation}
are of size $2N\times 2N$ and $2\times 2$-block-tridiagonal.
The single-particle components are given by
\begin{align}
    \hat{H}_{\text{d}} (k) 
    &= \left( - \mu - 2 t \cos k \right) \hat{\sigma}^0 + \alpha \sin k \hat{\sigma}^y + h \hat{\sigma}^z \\
    \begin{split}
        \hat{H}_{\text{od}} (k) 
        &= \left( - t - 2 t' \cos k \right) \hat{\sigma}^0 \\
        &\Peq -\frac{i}{2} \left( \alpha + \alpha' \cos k \right) \hat{\sigma}^x + \frac{\alpha'}{2} \sin k \hat{\sigma}^y
    \end{split}
\end{align}
while the gap components by
\begin{align}
    \begin{split}
        \hat{\Delta}_{\text{d}} (k) 
        &= \left[ \left( \Delta_s + \Delta_t \cos k \right) \hat{\sigma}^0 - \Delta_p \sin k \hat{\sigma}^y \right. \\
        &\Peq \left. -i \Delta_{p'} \sin k \hat{\sigma}^x \right] (i \hat{\sigma}^y) 
    \end{split} \\
    \hat{\Delta}_{\text{od}} (k) 
    &= \left[ \Delta_t \hat{\sigma}^0 - i \Delta_p \hat{\sigma}^x - \Delta_{p'} \hat{\sigma}^y \right] (i \hat{\sigma}^y),
\end{align}
where the labels \emph{d} and \emph{od} indicate that the components are diagonal and off-diagonal, respectively.

\section{Precession Terms and Symmetrization of the Current Operators}
\label{app:currents}

To derive the expressions for the spin and charge current operators, we generally follow the procedure briefly outlined in Sec.~\ref{sec:strip:micro}.
Starting from Eq.~\eqref{eq:strip:heisenberg}, assuming an infinitely extended system, we perform a Fourier transform along both the $x$ and $y$ direction.
Expanding for small $\Q$ up to linear order allows us to identify the terms proportional to $\Q$ as current terms, $\bs{J}_{\Q}^{\nu}$, and all other terms as precession terms, $P_{\Q}^{\nu}$.
For these additional precession terms appearing in the continuity equation for the spin currents, we find a generalization of the results in \cite{vorontsov:2008, erlingsson:2005} including also next-nearest neighbor contributions
\begin{align}
    P_{\Q}^{x} 
    &= \frac{\alpha}{t} \left( J_{\Q}^{x, z} \right)^{\text{nn}} + \frac{\alpha'}{2 t'} \left( J_{\Q}^{x, z} \right)^{\text{nnn}} \\
	P_{\Q}^{y} 
	&= \frac{\alpha}{t} \left( J_{\Q}^{y, z} \right)^{\text{nn}} + \frac{\alpha'}{2 t'} \left( J_{\Q}^{y, z} \right)^{\text{nnn}} \\
	\begin{split}
    	P_{\Q}^{z} 
    	&= -\frac{\alpha}{t} \left( J_{\Q}^{x, x} + J_{\Q}^{y, y} \right)^{\text{nn}} \\
    	&\Peq - \frac{\alpha'}{2 t'} \left( J_{\Q}^{x, x} + J_{\Q}^{y, y} \right)^{\text{nnn}}.
	\end{split}
\end{align}
Note that the precession terms appear in the continuity equation [Eq.~\eqref{eq:strip:continuity_eq}] and are only stated for completeness.
They are not needed in the analysis of the current expressions.

Following Sec.~\ref{sec:strip:micro}, the charge and spin currents $\mathcal{J}^{x, \nu}(y)$ are computed according to Eq.~\eqref{eq:strip:gs_currents} and evaluated for $y=l$ and $y=l\!+\!1/2$, $1 <= l < N_y$.
However, for the visualization of the numerical simulations, we redefine the current density operators in a more symmetric way and distribute the off-diagonal contributions at $y=l\!+\!1/2$ evenly to the adjacent sites $l$ and $l+1$ such that the total current at $y=l$ is given by \cite{imai:2012}
\begin{equation}
    \mathcal{J}_{l}^{x, \nu} + \frac{1}{2}
	\begin{cases}
		\mathcal{J}_{1+\frac{1}{2}}^{x, \nu} & l = 1 \\
		\mathcal{J}_{l+\frac{1}{2}}^{x, \nu} + \mathcal{J}_{l-\frac{1}{2}}^{x, \nu} & l = 2, ..., N_y-1 \\
		\mathcal{J}_{N_y-\frac{1}{2}}^{x, \nu} & l = N_y.
	\end{cases}
\end{equation}

\end{document}